\documentstyle[prl,aps,multicol,epsfig]{revtex}

\begin{document}

\def\ra{\rangle}
\def\la{\langle}
\def\bege{\begin{equation}}
\def\ende{\end{equation}}
\def\begarr{\begin{eqnarray}}
\def\endarr{\end{eqnarray}}
\def\no{\noindent}\def\non{\nonumber}
\def\hi{\hangindent=45pt}
\def\v{\vskip 12pt}

%
%
\newcount\hourCount
\hourCount=\time \divide\hourCount by 60\newcount\minuteCount
\minuteCount=\time \multiply\hourCount by 60 \global\advance\minuteCount
by -\hourCount \divide\hourCount by
60\newcommand{\hour}{\number\hourCount}\ifnum\minuteCount<10
\newcommand{\minute}{0\number\minuteCount}\else
\newcommand{\minute}{\number\minuteCount}\fi%
%
\newcommand{\figref}[1]{Fig.\ \ref{#1}}
\newcommand{\ket}[1]{{\left|#1\right>}}
\newcommand{\bra}[1]{{\left<#1\right|}}
\newcommand{\etal}{\textit{et al.}}
%
%

\draft

\title{
Narrowing of EIT resonance in a Doppler Broadened Medium
}

\author{
Ali~Javan,$^{1,2}$
Olga~Kocharovskaya,$^{3,4}$
Hwang~Lee,$^{4}$\footnote{
Present address: Jet Propulsion Laboratory,
MS 126-347,\\
California Institute of Technology,
Pasadena, CA~91109}
and
Marlan~O.~Scully$^{2,4}$
}

\address{$^1$Department of Physics,
Massachusetts Institute of Technology,
      Cambridge, MA~02139-4307\\
$^2$ Max-Planck-Institut f\"{u}r Quantenoptik,
        D-85748 Garching, Germany \\
$^3$ Institute of Applied Physics, RAS,
        Nizhny Novgorod
603120, Russia \\
$^4$ Department of Physics, Texas A\&M University,
College Station, TX~77843-4242
}
\date{October 30, 2001}

\maketitle

\begin{abstract}
We derive an
analytic expression for the linewidth of EIT resonance
in a Doppler broadened system.
It is shown here that for relatively low intensity
of the driving field the EIT linewidth is proportional
to the square root of
intensity and is independent of the Doppler width,
similar to the laser induced line narrowing effect
by Feld and Javan. 
In the limit of high intensity we recover the usual
power broadening case where EIT linewidth is proportional
to the intensity and inversely proportional to the Doppler
width.
\end{abstract}


\begin{multicols}{2}

Due to the Doppler effect
the atoms in a gas see the radiation field with shifted frequency.
Hence the macroscopic polarization representing medium's response
to the radiation,
needs to be averaged over the frequency
distribution determined by velocity distribution of the atoms.
By and large,
all sorts of phenomena in gas laser are related to Doppler
broadening \cite{ssl74} and
it is also the origin of the famous hole burning \cite{bennett62}.
and Lamb dip \cite{szoke63,mcfar63}.
It was more than
thirty years ago that laser induced line narrowing effect
in a three-level
Doppler broadened system was discovered by Feld and Javan \cite{feld69}.
Notably Feld and Javan found the
spectral width of the narrow line to be linearly
proportional to the driving
field Rabi frequency. Various aspects of this effect have been
investigated
\cite{popova70,hansch70,feldman72}.

The interest to the narrow nonabsorption
resonances imposed on the Doppler profile has been resumed recently in
a connection with the
Electromagnetically Induced Transparency (EIT) 
experiments which
have produced an ultra-slow
light propagation~\cite{hau99,kash99,budker99}
with spatial compression
(group velocity less than 10's m/sec) and
have made it possible to enhance
nonlinear optical processes by orders of
magnitude \cite{hakuta91,hemmer95,jain96,zibrov99}.

Steepness of the dispersion
function with respect to frequency plays the key role for
the small group velocity of light,
and is directly related to the transmission
width~\cite{harris92,xiao95,schmidt96}.
Hence the behavior of the
transmission linewidth in terms of experimental parameters is of a
great deal of interest.
In high resolution spectroscopy and high
precision magnetometry based on a narrow EIT
line \cite{mos92,mfl94,brandt97,lukin97,budker98,nagel98}
the experiments
are usually carried out with atomic cell
configurations so that the effect
of Doppler broadening on EIT is also an important concern for
the performance of the devices.

Doppler broadening effects in EIT and lasing without inversion
(LWI) have been studied in a number of works
\cite {arimondo96,li95,kara95,vemuri96,wang95}. Most of these
works focused on the possibilities of absorption cancellation
and preferable field configurations (co-propagation of probe and drive
lasers in folded
schemes, counter-propagation in cascade schemes).
In the limit of the
vanishing probe field and under the assumption
that all atoms were trapped to the dark state
it was found that the power broadening of EIT line takes
place:
$\Gamma_{EIT}=\Omega^2 / W_D$
(where $\Omega$ is the Rabi frequency of the driving field
and $W_D$ is Doppler linewidth),
which is
similar to the well-known result for the homogeneously
broadened system:
$\Gamma_{EIT}=\Omega^2 / \gamma$
(where $\gamma$ is a homogeneous linewidth).
This dependence was experimentally verified in \cite{kash99}.
In the
limit of relatively low probe field intensity,
$\alpha \ll (\gamma/W_D)\Omega$,
and under the same assumption of full coherent trapping
(i.e.\ neglecting by the two-photon coherence decay)
it leads to the following result
for EIT line width:
$\Gamma_{EIT}= \alpha \Omega / \gamma$,
where $\alpha$ is the
Rabi frequency 
of the probe field \cite{taich00}.

In this paper,
we find  an explicit expression for the linewidth of
EIT resonance in
a Doppler broadened three-level system
in the linear approximation with
respect to probe field
taking into account finite decay time of
low-frequency coherence.
In the limit of very large intensity it is reduced
to the power broadening case.
However, for the intermediate range of
intensities 
the coherent population trapping is velocity selective,
i.e., it occurs only for those atoms whose frequencies are close
to the resonance with a driving field.
In this case
we find that the width of EIT resonance
is proportional to the Rabi frequency of the driving field
(similar to result by Feld and Javan \cite{feld69})
and to the square root of the ratio of
the relaxation times of the coherence at the two-photon
(low-frequency) and population difference at one-photon (optical)
transitions:

\begin{eqnarray}
\Gamma_{EIT}
&\Longrightarrow&  ~~\sqrt {2\gamma_{bc} \over \gamma} ~\Omega .
\end{eqnarray}

\no
This regime corresponds to the narrowest possible
EIT line-width and therefore it is very favorable
for realization of the
efficient EIT-based nonlinear transformations and light storage.

Let us
consider the closed atomic model scheme depicted in Fig.~1.
In this
three-level $\Lambda$ scheme one of the two lower-levels is
coupled
to the upper level ($a\rightarrow c$) by a coherent drive laser
and the transition
$a\rightarrow b$ is probed by a weak coherent field.
The atomic decays are
confined among the given levels.
Note that such a model gives a
description almost equivalent to the one for an open system
in which atoms
decay (out of the interaction region) with the 
rate $\gamma_{bc}$,
and atoms are coming into the interaction region 
with equally populated lower levels.
Detailed comparison of our model with the open system will be
published elsewhere.

\begin{figure}[htb]
\epsfysize=3.5cm
\centerline{\epsfbox{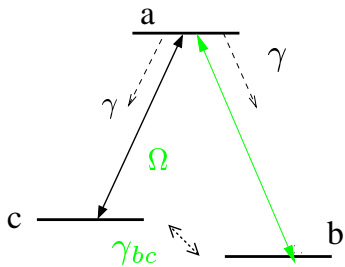}}~
\caption{\label{fig-1}
Three-level model scheme.
The upper level $a$ decays to $b$ and $c$ with decay rate $\gamma$.
The relaxation rate between levels $b$ and
$c$ is denoted
as $\gamma_{bc}$, which is assumed to be small
compared to $\gamma$.}
\end{figure}

If the system is Doppler broadened, the
susceptibility should be averaged over the entire velocity distribution
such
that \cite{ssl74}

\begin{equation}
\chi     =        \int d(kv)~f(kv) ~\eta
\left\{ {\rho_{ab}(kv) \over \alpha} \right\},  \label{j-1}
\end{equation}

\no
where $k$ is the wave number of the probe field,
$f(kv)$ is
the velocity distribution function,
$\rho_{ab}(kv)$ is the coherence
between states $a$ and $b$ induced by radiation fields,
${\eta \equiv (3/8\pi) N \gamma \lambda^3}$,
$N$ is the atomic density, 
and $\lambda$ is the wavelength.
For a stationary atom,
in the first order of the probe field,
$\rho_{ab}$ can be written as
\begin{equation}
\rho_{ab} 
=  {-i\alpha \over \Gamma_{ab}\Gamma_{cb}
+\Omega^2}        \Big[ \Gamma_{cb} (\rho_{aa}^{(0)}-\rho_{bb}^{(0)})
+ {\Omega^2 \over \Gamma_{ca}}        (\rho_{cc}^{(0)}-\rho_{aa}^{(0)})
\Big] ,  \label{j-2}
\end{equation}

\no
where
$\rho^{(0)}_{ii}$'s are the zeroth
order populations
(in the probe field) and
${\Gamma_{ij} \equiv \gamma_{ij} + i \Delta_{ij}}$
with the off-diagonal decay rates $\gamma_{ij}$ given
by
$\gamma_{ab} = \gamma_{ac} =(\gamma + \gamma^{\prime} + \gamma_{bc})/2$,
$\gamma_{cb} = \gamma_{bc}$.
$\Delta_{ij}$'s are defined
as $\Delta_{ab}=\omega_{ab}-\nu \equiv
\Delta$, $\Delta_{ac}=\omega_{ac}-\nu_0$,
and
$\Delta_{cb}=\Delta_{ab}-\Delta_{ac}$,
where $\nu$ and $\nu_0$ are the
frequencies of the probe and drive fields, respectively.

In the present analysis we
use the following assumptions:
1) The decay rates in the transitions
$a \rightarrow b$ ($\gamma$) and
$a \rightarrow c$ ($\gamma'$) are assumed to
be same ($\gamma$) and defined by spontaneous emission,
which is typically the case for the dilute gases.
2) The decay rates of population difference and
coherence at the low-frequency transition $b \leftrightarrow c$
are the same ($\gamma_{bc}$),
which is typically the case when this decay is determined by the time of
flight through the interaction region.
3) The probe field is weak such that
the first order analysis is valid.
4) The driving field is on resonance for
a stationary atom: $\omega_{ac}=\nu_0$.
5) The probe field and driving field propagate in the
same direction,
and the frequency difference between the transitions
$a \rightarrow b$ and $a \rightarrow c$ is small enough such that the
residual Doppler shift,
$(k-k^{\prime})v$, can be ignored.
6) The EIT condition for the
homogeneously broadened system
($\Omega^2 \gg \gamma \gamma_{bc}$)
is valid.
7) The inhomogeneous linewidth ($W_D$) is large enough such that
$W_D \gg \gamma, \Omega$.

Under thses assumptions the atomic
populations $\rho^{0}_{ii}$ can be written as
\begin{eqnarray}
\rho_{aa}^{(0)} &=& { 2
\gamma_{bc} \Omega^2 \over 2 D} ,\qquad
\rho_{cc}^{(0)} = { 4 \gamma X \gamma_{bc} + 2
\gamma_{bc} \Omega^2  \over 2D},
\non \\ 
\rho_{bb}^{(0)} &=& { 4 \gamma X
\gamma_{bc} + 2 \gamma_{bc} \Omega^2 + 2 \Omega^2 \gamma \over 2D },
\label{h4c}
\end{eqnarray}
where
$
X    =       [\gamma^2 + (kv)^2] / 2 \gamma
$, and
$
D   =  4 \gamma X \gamma_{bc} + 3 \gamma_{bc}
\Omega^2         + \Omega^2 \gamma
$.
Then, for an atom with its velocity $v$,
the off-diagonal element of the density matrix $\rho_{ab}(kv)$
is found as
\begin{eqnarray}
\rho_{ab} 
&=&        { i
\alpha \over Y }        {1 \over 2 D}        \Big[%
\Gamma_{c b}          (4 \gamma X \gamma_{bc}
+ 2 \Omega^2 \gamma)
  -
\frac{\Omega^2  4 \gamma X \gamma_{bc} }           { \gamma +
\gamma_{bc}/2 + i k v }        \Big] ,  \label{j-4}
\end{eqnarray}%
where
$Y= (\gamma +\gamma_{bc}/2 +i \Delta + i k
v)(\gamma_{bc}+i\Delta) + \Omega^2$.

Doppler broadening is usually modeled by convolution of a given
function over a Maxwell-Boltzmann velocity distribution.
Due to the
complexity in the integration with a Gaussian distribution,
however,
explanations of the obtained results usually rely on numerical
analysis \cite{li95,kara95,vemuri96,wang95}.
In order to obtain a simple
expression of the linewidth,
we approximate the usual Gaussian distribution
with a Lorentzian function;
this leads to a rather simple form of
inhomogeneously broadened susceptibility
with which detailed analysis is
possible.

If we use a Lorentzian profile as the velocity
distribution function $f(kv)$ with 
full width half maximum (FWHM)
$2W_D$ such that $f(kv) = (1/\pi) W_D/[W_D^2 + (kv)^2]$,
the
Eq.~(\ref{j-1}) can be evaluated by the contour integration in
the complex
plane which contains two poles in the lower half plane, viz.,
$kv = - i W_D$ and
$kv = -i \sqrt{\Omega^2 \gamma/ 2 \gamma_{bc}}$.
After
straight forward calculation of the contributions from the two poles,
one can
find the complex susceptibility.
In particular, the minimum absorption at
the line center is obtained as%
\begin{equation}
\chi^{\prime \prime}
(\Delta=0)     =        {\eta \gamma_{bc} \over \gamma_{bc} W_D +
\Omega^2}        \left[          {\sqrt{x} \over 1 + \sqrt{x}}
\right].   \label{min} \end{equation}%
\no
where $x = {\Omega^2 \gamma / 2 \gamma_{bc} W_D^2}$.
We note that, as long as $\Omega^2 \gg \gamma \gamma_{bc}$,
the expression is vanishingly small
as $\eta \sqrt{x}/W_D$ when $x \ll 1$,
and also as $\eta \gamma/W_D^2$ when $x \gg 1$, so
that the EIT (i.e. strong suppression of absorption in the presence of
driving field at $\Delta=0$)
is preserved.
The maximum of $\chi^{\prime \prime}$, on the other hand, can be
found as
$\chi^{\prime \prime}_{\rm max} \approx \eta/W_D$
at $\Delta \approx \pm \Omega$.

Since the absorption at the line center is negligibly
small given by (\ref{min}),
we evaluate $\Delta$ which
defines $\Gamma_{EIT}$ as
$\chi^{\prime \prime} (\Delta=\Gamma_{EIT}) = \eta/2W_D$.
The half width of the EIT
resonance ($\Gamma_{EIT}$)
is, then, obtained
as

\begin{equation}
\Gamma_{EIT}^2={\gamma_{bc} \over \gamma}
~\Omega^2(1+x)\left[1+ \left\{ 1+ {4x \over (1+x)^2 }\right\}^{1/2}\right]
.
\label{j-9}
\end{equation}
This is the main result of the present
report.
Here we can see the two extreme cases,
namely,
\begin{mathletters}
\begin{eqnarray}
\Gamma_{EIT}
&\Longrightarrow&        ~~\sqrt {2\gamma_{bc} \over \gamma} ~\Omega
\qquad        (x \ll 1) ,   \label{fj-limit} \\    &\Longrightarrow&
~~~{\Omega^2 \over W_D}        \qquad \quad        (x \gg 1) .
\label{op-limit}
\end{eqnarray}
\end{mathletters}
Note that the range of
$x$ is:
$(\gamma/W_D)^2 \ll x \ll \gamma/\gamma_{bc}$.
In the expression
(\ref{fj-limit}) corresponding to the limit $x \ll 1$,
the linewidth of EIT
is linearly proportional to $\Omega$,
the Rabi frequency of the driving
field
(i.e., to the square root of the driving field intensity) and
it is
independent of the Doppler width $W_D$.

Similar linear dependence of the
linewidth on the Rabi frequency was previously obtained
in Ref.\ \cite{feld69}.
The earlier work \cite{feld69} dealt with
a laser gain system where the weak
transitions between the lasing levels were used.
The decays out of the
lasing levels were the main relaxation mechanisms
while the spontaneous decays between levels were
not taken into account.
These so-called open
systems have the relaxation of low-frequency coherence
($\gamma_{bc}$) the
same order of magnitude as the relaxation of population
difference at the optical transitions ($\gamma$),
i.e., $\gamma_{bc} \approx \gamma$.
In this
case we have
$x \approx \Omega^2/2W_D^2$,
the Eq.~(\ref{fj-limit}) takes a form:
$\Gamma_{EIT} \approx \Omega$.
Since $\Omega \ll W_D$, the linewidth, in turn,
is much smaller than $W_D$.
This limit fully
corresponded to experimental conditions of Ref.\cite{feld69}.

In the limit
$x\gg 1$ (corresponding to small $\gamma_{bc}$ 
or a strong driving field)
$\Gamma_{EIT}$
is proportional to intensity, $\Omega^2$,
and inversely proportional to $W_D$.
Many recent EIT experiments
were performed in alkali vapors where the two-photon
coherence, ($\rho_{bc}$)
was built among the hyperfine levels of the
ground state.
In these systems the low-frequency coherence relaxation time is
determined by the time of flight of the atom through
the interaction region,
and it is large as compared to the life time 
of the excited optical state.

In Fig.~2,
we plot the EIT linewidth as a
function of the Rabi frequency of the driving field.
We note
that
$\Gamma_{EIT} \geq \Omega \sqrt{2\gamma_{bc}/\gamma}$
for any value of
$\Omega$.
Apparently, smaller ratio $\gamma_{bc}/\gamma$ leads to
smaller EIT width at $x \ll 1$, and
to smaller value of $\Omega$ at which
the linear dependence of $\Omega$ in $\Gamma_{EIT}$
($\Gamma_{EIT} \propto \Omega$) changes to quadratic
dependence ($\Gamma_{EIT} \propto \Omega^2$).
Both in $x \ll 1$ and $x \gg 1$ limits,
for a given value of intensity,
the width of EIT resonance in the
inhomogeneously broadened medium is smaller than in homogeneously
broadened medium with the same homegeneous line width at resonant driving.
In the limit $x \gg 1$ this fact was outlined earlier in
\cite{taich00}.

\begin{figure}[htb]
\epsfysize=5cm
\centerline{\epsfbox{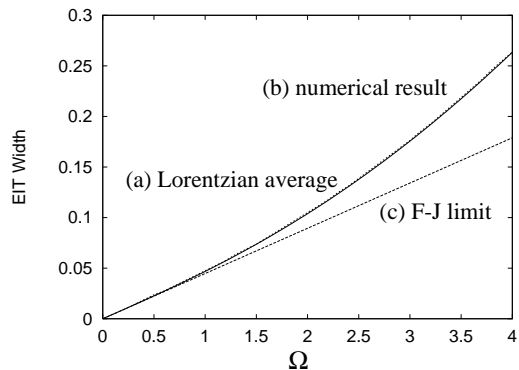}}
\caption{\label{fig-2}
EIT linewidth (in unit of $\gamma$) as a function of $\Omega$
(also in unit of $\gamma$), with Doppler width $2W_D=100\gamma$
and
$\gamma_{bc} = 10^{-3} \gamma$.
The plot (a) of Eq.~(\ref{j-9})
by averaging over the Lorentzian distribution function
(solid line) is
almost indistinguishable to that (b) of
the numerical result made by Gaussian integration (dotted line).
(c) F-J limit denotes the value of
$\Omega\sqrt{2\gamma_{bc}/\gamma}$.}
\end{figure}

This line narrowing effect has a simple physical
explanation.
Namely, it is due to the reduced power broadening for the
off-resonant atoms.
At the same time it is worth to note that the width of
EIT resonance in Doppler broadened system never can be reduced beyond the
ultimate limit defined by low-frequency coherence decay time:
$\Gamma_{EIT} \ge \gamma_{bc}$.
It reaches this limit when EIT sets in
with $\Omega^2  \ge \gamma \gamma_{bc}$
independently if the optical line is
homegeneously or inhomogeneously broadened.
In the case $x \gg 1$ EIT line
width exceeds this minimum value 
at least by the factor $W_D/\gamma$.

The physical meaning of the parameter $x$ can
be understood in the following way:
First, let us suppose the system
is homogeneously broadened.
The optical pumping rate from the level $c$ to
$b$ is $\Omega^2 / \gamma$
for the resonant driving field.
In order to have a
complete coherent optical pumping in the case of
resonant driving this
rate should be much bigger than the pumping rate from $b$ to $c$:
$\Omega^2/ \gamma \gg \gamma_{bc}$.
This means that the driving field should be
sufficiently strong:
$\Omega^2 \gg \Omega_{hom}^2 \equiv
\gamma_{bc}\gamma$.
For atoms with velocity $v$, then, the optical pumping
rate is
$\Omega^2 \gamma / [\gamma^2 + (kv)^2]$.
Then, in order to have a complete
coherent optical pumping in a Doppler broadened system we need to
require $\Omega^2 \gamma / (\gamma^2 + W_D^2) \gg \gamma_{bc}$,
which
corresponds to $x \gg 1$,
i.e., $\Omega^2 \gg \Omega^2_{inhom} \equiv 2\gamma_{bc}W_D^2/\gamma$.
Hence, the parameter $x$ represents the degree of
optical pumping from the level $c$ to $b$ within the
inhomogeneous line
width ($x = \Omega^2/\Omega^2_{inhom}$).

With a notion of the effective width $\delta_{\rm eff}$,
the width of EIT resonance can always be regarded as

\bege
\Gamma_{EIT} \sim {\Omega^2 \over
\delta_{\rm eff} },
\label{del-eff}
\ende

\no
which is equivalent to the EIT
linewidth for the homogeneously broadened medium (where
$\Gamma_{EIT}=\Omega^2/\gamma$).
The effective width $\delta_{\rm eff}$ is
defined as the magnitude of the maximum detuning for which atoms are
optically pumped into the level $b$
(and hence can interact with a probe
field) for a fixed value of $\Omega$.

For $\Omega_{hom} \ll \Omega \ll
\Omega_{inhom}$,
$\delta_{\rm eff}$ can be estimated by $\Omega^2 (\gamma
/\delta_{\rm eff}^2) \sim \gamma_{bc}$,
yielding $\delta_{\rm eff} \sim
\sqrt{\Omega^2 \gamma /\gamma_{bc}}$.
Therefore, an increase of intensity of the
driving field makes the number of trapped atoms increased,
which results,
according to Eq.~(\ref{del-eff}), in the
linear dependence of EIT resonance
width:
$\Gamma_{EIT} \sim {\Omega \sqrt{\gamma_{bc}/\gamma}}$
[see,
Eq.~(\ref{fj-limit})].
When $\Omega \gg \Omega_{inhom}$ the number of
optically pumped atoms is not increased further (since all of them are
already optically pumped into the level $b$),
so that $\delta_{\rm eff} \sim W_D$
yielding
$\Gamma_{EIT} = \Omega^2/W_D$.

It is worth to note that obtained results can be used for description
of EIT experiments not only in gaseous media with Doppler broadening
but also in solids with the long lived spin coherence,
for example, in
rare-earth ions doped crystals at low temperature\cite{hemmer01}
when inhomogeneous line
broadening of optical transitions plays a major role
while inohomogeneous
broadening of the spin transitions is negligible.
On the other hand, they
are not directly applicable for EIT experiments involving
a buffer gas in
a cell or paraffin coating
since collisions of the operating atoms
with the buffer gas or wells can essentially disturb
the Doppler velocity
distribution.

The authors wish to thank C.\ J. Bednar,
B.\ G.\ Englert, M.\ S.\ Feld, M.\ D.\ Lukin, A.\ B.\ Matsko,
Yu.\ Rostovtsev, V.\ L.\ Velichansky, A.\ S.\ Zibrov for helpful
and stimulating discussions.
This work was supported by the Office of Naval Research, 
Defense Advanced Research Projects Agency, 
Texas Advanced Research Program, and 
the Air Force Research Laboratories.

\end{multicols}

\end{document}